\documentstyle[11pt,aasms4]{article}

\newcommand{\vx}{{\bf x}}
\newcommand{\vxp}{{\bf x^\prime}}
\newcommand{\vv}{{\bf v}}
\newcommand{\vk}{{\bf k}}
\newcommand{\divg}{{\vec\nabla}\cdot}
\newcommand{\vpred}{{\bf v}_{\mathrm{pred}}}
\newcommand{\vtrue}{{\bf v}_{\mathrm{true}}}
\newcommand{\vlin}{{\bf v}_{\mathrm{lin}}}
\newcommand{\vper}{{\bf v}_{\mathrm{per}}}
\newcommand{\vavg}{\bar{{\bf v}}}
\newcommand{\vsig}{\mbox{\boldmath $\sigma$}}
\newcommand{\vtruei}{{\bf v}_{\mathrm{true},\mathit{i}}}
\newcommand{\vpredi}{{\bf v}_{\mathrm{pred},\mathit{i}}}
\newcommand{\hmpc}{h^{-1}\mathrm{Mpc}}

\begin{document}
 
\title{Biased Estimates of $\Omega$ from Comparing Smoothed Predicted
Velocity Fields to Unsmoothed Peculiar Velocity Measurements}

\author{Andreas A. Berlind, Vijay K. Narayanan \altaffilmark{1} and David H. 
Weinberg}
\affil{Department of Astronomy, The Ohio State University, Columbus, OH 43210;
aberlind,dhw@astronomy.ohio-state.edu}
\altaffiltext{1}{Present Address: Department of Astrophysical Sciences, 
Princeton University, Princeton, NJ 08544-1001; vijay@astro.princeton.edu}

\begin{abstract}

We show that a regression of unsmoothed peculiar velocity measurements
against peculiar velocities predicted from a smoothed galaxy density
field leads to a biased estimate of the cosmological density parameter 
$\Omega$, even when galaxies trace the underlying mass distribution and
galaxy positions and velocities are known perfectly.  The bias arises
because the errors in the predicted velocities are correlated with the
predicted velocities themselves.  We investigate this bias using 
cosmological N-body simulations and analytic arguments.  In linear
perturbation theory, for cold dark matter power spectra and Gaussian 
or top hat smoothing filters, the bias in $\Omega$ is always positive,
and its magnitude increases with increasing smoothing scale.
This linear calculation reproduces the N-body results 
for Gaussian smoothing radii $R_s \ga 10\hmpc$, while non-linear
effects lower the bias on smaller smoothing scales, and for $R_s \la 3\hmpc$ 
$\Omega$ is underestimated rather than overestimated.  
The net bias in $\Omega$ for a given smoothing filter depends on the
underlying cosmological model.  The effect on current estimates of $\Omega$ 
from velocity-velocity comparisons is probably small relative to other
uncertainties, but taking full advantage of the statistical precision of 
future peculiar velocity data sets will require either equal smoothing of
the predicted and measured velocity fields or careful accounting for the
biases discussed here. 

\end{abstract}


\section{Introduction}

One of the most popular approaches to constraining the mass density 
parameter $\Omega$, the ratio of the average matter density to the critical
density,  is based on comparisons between the galaxy density field mapped 
by redshift surveys and the galaxy peculiar velocity field inferred from 
distance-indicator surveys (see the review by \cite{strauss95}).  While 
the numerous implementations of this approach differ in many details, they 
are all motivated by the linear theory formula for the peculiar velocity 
field,
\begin{equation}
\vv(\vx) = {H_0 f(\Omega) \over 4\pi} \int \delta(\vxp) 
	   {(\vxp-\vx) \over |\vxp - \vx|^3} d^3x^\prime ~,
\label{eqn:vlin}
\end{equation}
or its divergence 
\begin{equation}
\divg\vv(\vx) = - a_0 H_0 f(\Omega) \delta(\vx) ~,
\label{eqn:divv}
\end{equation}
where $\delta(\vx) \equiv \rho(\vx)/\bar{\rho} - 1$ is the mass density
contrast, $f(\Omega) \approx \Omega^{0.6}$, $H_0$ is the Hubble parameter,
and $a_0$ is the present value of the expansion factor 
(\cite{peebles80}).\footnote{Because galaxy distances are inferred
from their redshifts via Hubble's law, uncertainties in $H_0$ and $a_0$
do not introduce any uncertainy in peculiar velocity predictions;
if one adopts km s$^{-1}$ distance units in place of Mpc, then $H_0$ and $a_0$ 
do not appear in equation~(\ref{eqn:vlin}) or~(\ref{eqn:divv}).}
``Velocity-velocity'' comparisons start from the observed galaxy density
field, predict peculiar velocities via equation~(\ref{eqn:vlin}) or some
non-linear generalization of it, and compare to estimated peculiar velocities
(e.g., \cite{kaiser91}; \cite{strauss95}; \cite{davis96}; \cite{willick97},
1998; \cite{blakeslee99}).
``Density-density'' comparisons start from the observed radial peculiar 
velocity field, infer the 3-dimensional velocity field using the POTENT 
method of Bertschinger \& Dekel (1989), and compare the velocity divergence 
to the observed galaxy density field using equation~(\ref{eqn:divv}) or a 
non-linear generalization of it (e.g., \cite{dekel93}; \cite{hudson95}; 
\cite{sigad98}; \cite{dekel99}).  Because the radial velocity field must be 
smoothed before computing the 3-dimensional velocity field via POTENT, 
density-density comparisons in practice always compare the smoothed galaxy 
density field to predictions derived from the smoothed peculiar velocity field.  
Velocity-velocity comparisons, on the other hand, usually smooth the galaxy 
density field to suppress non-linear effects and shot noise, but compare the 
velocity predictions from these smoothed density fields directly to the 
estimated peculiar velocities of individual galaxies or groups. (The spherical
harmonic analysis of Davis et al.\ 1996 is an important exception in this regard.)

The avoidance of smoothing the data is often seen as an advantage of the 
velocity-velocity approach, since smoothing a noisy estimated velocity 
field can introduce statistical biases that are difficult to remove.  
However, in this paper we show that comparing smoothed velocity 
predictions to unsmoothed velocity measurements generally leads to biased 
estimates of $f(\Omega)$, even when the galaxy positions and velocities
are known perfectly.  The reason for this bias is fairly simple:
the errors in the predicted velocities are correlated with the predicted 
velocities themselves, violating the conventional assumption that an individual 
galaxy's velocity can be modeled as a ``large scale'' contribution predicted
from the smoothed density field plus an {\it uncorrelated} ``small scale'' 
contribution.

Galaxy redshift surveys map the galaxy density field $\delta_g(\vx)$
rather than the mass density field $\delta(\vx)$, so inferences
from velocity-velocity and density-density comparisons often assume a
linear relation between galaxy and mass density contrasts,  
$\delta_g(\vx)=b\delta(\vx)$, and therefore
constrain the quantity $\beta \equiv f(\Omega)/b$ rather than $f(\Omega)$
itself.  The results reported in this paper emerged from a more 
general investigation of the effects of complex galaxy formation models 
on estimates of $\beta$ (\cite{berlind99}).  However, the 
statistical bias in $f(\Omega)$ that we find applies even when galaxies trace 
mass exactly, so here we focus on this simpler case.  (Throughout this paper
we use the term ``bias'' to refer to systematic statistical errors
rather than the relation between the distributions of galaxies and mass.)
We further restrict our investigation to the case in which galaxy positions 
and velocities are known perfectly, ignoring the additional complications 
that arise in analyses of observational data.

\section{Results}

We have carried out N-body simulations of three different cosmological
models, all based on inflation and cold dark matter (CDM).
The first is an $\Omega=1$,
$h=0.5$ model ($h \equiv H_0/100\;{\rm km}\;{\rm s}^{-1}\;{\rm Mpc}^{-1}$),
with a tilted power spectrum of density
fluctuations designed to satisfy both COBE and cluster normalization
constraints.  
The cluster constraint requires $\sigma_8\approx0.55$ (\cite{white93}),
where $\sigma_8$ is the rms linear density fluctuation in spheres of radius 
$8h^{-1}$Mpc.  Matching the COBE-DMR constraint and $\sigma_8=0.55$ with
$h=0.5$ requires an inflationary spectral index $n=0.803$ if one incorporates
the standard inflationary prediction for gravitational wave contributions to
the COBE anisotropies (see \cite{cole97} and references therein).  
The other two models have $\Omega=0.2$ and $0.4$, with
a power spectrum shape parameter $\Gamma=0.25$ (in the parameterization of
\cite{efstathiou92}) and cluster-normalized 
fluctuation amplitude $\sigma_8=0.55 \Omega^{-0.6}$.  We ran 
four independent simulations for each of the three cosmological models, and the
results we show below are averaged over these four simulations. All
simulations were run with a particle-mesh 
(PM) N-body code written by C. Park, which is described and tested by 
\cite{park90}.  Each simulation uses a $400^3$ force mesh to follow the 
gravitational evolution of $200^3$ particles in a periodic cube $400 h^{-1}$Mpc
on a side, starting at $z=23$ and advancing to $z=0$ in 46 steps of
equal expansion factor $a$.
We form the mass density field by cloud-in-cell (CIC) binning the evolved
mass distribution onto a $200^3$ grid.  We smooth this density field with
a Gaussian filter of radius $R_s$ and derive the linear-theory predicted
velocity field using equation~(\ref{eqn:vlin}).  Finally, we linearly
interpolate this velocity field to the galaxy positions to derive 
predicted galaxy peculiar velocities $\vpred$.

Figure 1 compares the true velocities of particles ($\vtrue$) from one of 
the $\Omega=1$ simulations to the velocities predicted ($\vpred$) by 
equation~(\ref{eqn:vlin}) from the mass density field smoothed with Gaussian 
filters of radius $R_s=3,5,10,$ and $15\hmpc$ (panels a-d, respectively).
The points in Figure 1 show one Cartesian component of the particles' 
velocities.  If we make the assumption, common to most velocity-velocity 
comparison schemes, that each galaxy's velocity consists of a large scale
contribution predicted from the density field plus an uncorrelated small scale
contribution, then the best-fit slope of the $\vtrue - \vpred$ relation should
yield the parameter $f(\Omega)$, in this case $f(\Omega)=1$, with the scatter
about this line yielding the dispersion of the small scale contribution.  
However, it is clear from Figure 1 that this slope increases systematically
with increasing $R_s$.  (We note that the best-fit line, which minimizes
$\sum|\vtrue-\vpred|^2$, is shallower than the line one would naively draw
through these data points by eye, since it is vertical scatter rather than
perpendicular scatter that must be minimized.)  

The filled points in Figure 2 show the estimated $f(\Omega)$ as a function
of $R_s$ for the $\Omega=1$ (circles) and $\Omega=0.2$ (squares) cosmological
models.  The solid lines show the true value of $f(\Omega)$.
In both cases, the estimated value of $f(\Omega)$ is quite sensitive to the 
smoothing scale: it is slightly underestimated at small scales, but increasingly
overestimated at large scales.  The $\Omega=0.4$ model yields similar results,
so we do not plot it separately.  We also investigated $\Omega=1$ simulations
with a factor of two lower force resolution ($200^3$ force mesh instead of
$400^3$) and found identical results, so even at small smoothing scales our
results are not affected by the simulations' limited gravitational resolution.
The breakdown of linear theory at small scales is not surprising; however,
the systematic failure of this method at large smoothing scales has not, to 
our knowledge, been previously discussed.  The dependence of the estimated
$f(\Omega)$ on the smoothing scale used for velocity predictions is our
principal result.

We can understand the origin of the large scale bias in $f(\Omega)$ by
considering the case in which galaxy peculiar velocities are given exactly by 
linear theory.  In this case,
\begin{equation}
\vtrue(\vx) = (2\pi)^{3/2}H_0f(\Omega)\int e^{i \vk \cdot \vx} 
         \frac{i\delta_{\vk} \vk}{|\vk|^2} d\vk ~,
\label{eqn:vtrue}
\end{equation}
where $\delta_{\vk}$ are the Fourier modes of the density field and the
integral extends over all of $\vk$-space.  Predicted velocities,
however, are estimated from the density field smoothed with a window
function $W(r)$ of characteristic scale $R_s$.  Therefore, 
\begin{equation}
\vpred(\vx) = (2\pi)^{3/2}H_0f(\Omega)\int \widetilde{W}(kR_s) 
	 e^{i \vk \cdot \vx} 
         \frac{i\delta_{\vk} \vk}{|\vk|^2} d\vk ~,
\label{eqn:vpred}
\end{equation}
where $\widetilde{W}(kR_s)$ is the Fourier transform of the window function.
The error in the predicted velocity of a galaxy at position $\vx$ is therefore,
\begin{equation}
\Delta \vv(\vx) = \vtrue-\vpred = (2\pi)^{3/2}H_0f(\Omega)
	     \int [1-\widetilde{W}(kR_s)] 
             e^{i \vk \cdot \vx} \frac{i\delta_{\vk} \vk}{|\vk|^2} d\vk ~.
\label{eqn:dv}
\end{equation}
Note that in equation~(\ref{eqn:vpred}) we have defined $\vpred$ to be the
velocity that would be predicted assuming the correct value of $\Omega$.
In practice, since we do not know the value of $f(\Omega)$ beforehand, we 
derive its value from the slope of the $\vtrue$ vs. $f^{-1}\vpred$ relation 
(this is equivalent to assuming $\Omega=1$ when computing $\vpred$).
 
If $\Delta \vv$ were uncorrelated with $\vpred$, then the slope of the $\vtrue$ 
vs. $f^{-1}\vpred$ relation would be an unbiased estimator of $f(\Omega)$.
However, if $\Delta \vv$ is positively correlated with $\vpred$, then the
slope of the relation is no longer $f(\Omega)$, since points preferentially
scatter above the line for positive $\vpred$ and below the line for negative 
$\vpred$.  This steepening of the $\vtrue - \vpred$ relation is just the
behavior seen in Figure 1.  Equations~(\ref{eqn:vpred}) and ~(\ref{eqn:dv})
show that $\Delta \vv$ and $\vpred$ will be correlated as long as some Fourier
modes contribute to both integrals, which happens for any smoothing function
other than a step function in $\vk$-space.  

We can quantitatively understand this bias by considering how $f(\Omega)$ is 
measured.  For an ensemble of $N$ points ($\vtruei$, $f^{-1}\vpredi$), the
slope of the best-fit line (assuming
$\langle\vtrue\rangle=\langle\vpred\rangle=0$) is
\begin{eqnarray}
\mathrm{slope} & = & \frac{\sum (f^{-1}\vtruei \cdot \vpredi)}
                          {\sum (f^{-2}\vpredi \cdot \vpredi)} \nonumber \\
               & = & f(\Omega)\frac{\frac{1}{N}\sum [(\vtruei-\vpredi) \cdot 
		     \vpredi
                     + (\vpredi \cdot \vpredi)]}
                          {\frac{1}{N} \sum (\vpredi \cdot \vpredi)} 
			  \nonumber \\
               & = & f(\Omega)\left[1 + \frac{\langle\Delta \vv \cdot 
					      \vpred\rangle}
                                   {\langle\vpred \cdot \vpred\rangle}\right] ~.
\label{eqn:slope}
\end{eqnarray}
Equation~(\ref{eqn:slope}) shows how a non-zero cross-correlation between 
$\Delta \vv$ and $\vpred$ changes the measured slope of the velocity-velocity
relation.  We can compute this effect in the linear regime for a given power 
spectrum of density fluctuations $P(k)$ and window function
$\widetilde{W}(kR_s)$.  Using equations~(\ref{eqn:vpred}) and ~(\ref{eqn:dv}) 
we have
\begin{eqnarray}
\frac{\langle\Delta \vv \cdot \vpred\rangle}{\langle\vpred \cdot \vpred\rangle} 
& = & \frac{\int_0^\infty \widetilde{W}(kR_s)[1-\widetilde{W}(kR_s)] P(k) dk}
           {\int_0^\infty \widetilde{W}^2(kR_s) P(k) dk} ~.
\label{eqn:dvvpred}
\end{eqnarray}
For Gaussian and top hat window functions and a range of CDM power spectra, we
find that the bias given by equation~(\ref{eqn:dvvpred}) is always positive
and is always an increasing function of $R_s$.  The dashed lines in Figure 2 
show the slope computed (from eqs.~\ref{eqn:slope} and \ref{eqn:dvvpred})
using the linear mass power spectra of the simulations and the same Gaussian
window functions that were used to measure $f(\Omega)$ (solid points).  The
striking similarity on large smoothing scales between the N-body data and this
linear theory calculation supports our conclusion that the large scale bias is
indeed caused by the cross-correlation between $\Delta \vv$ and $\vpred$,
which, in turn, is caused by the comparison of a smoothed prediction to 
unsmoothed data.

{}From equation~(\ref{eqn:dvvpred}) it is evident that the linear theory 
cross-correlation between $\Delta \vv$ and $\vpred$ will be equal to zero
if there is no smoothing at all, or if the smoothing function is a step
function in $\vk$-space, in which case the product
$\widetilde{W}(kR_s)[1-\widetilde{W}(kR_s)]$ is always equal 
to zero.  The open symbols in Figure 2 show results of a velocity-velocity 
analysis of the same simulations, with the linear theory velocities now
predicted from a density field smoothed with a sharp, low-pass $\vk$-space
filter.  Specifically, we set to zero all Fourier modes with 
$k>k_{\mathrm{cut}}$ and plot the new estimates of $f(\Omega)$ at the values
of $R_s$ for which a Gaussian filter falls to half its peak value at 
$k = k_{\mathrm{cut}}$ (i.e., $e^{-k_{\mathrm{cut}}^2 R_s^2/2} = 0.5$).  
Using the sharp $\vk$-space filter causes the bias to vanish completely on
large scales, yielding estimates of $f(\Omega)$ that are correct and
independent of smoothing length.  This result further supports our
interpretation of the cause of the large scale velocity-velocity bias.

Figure 2 shows that $f(\Omega)$ is underestimated at small scales in the
N-body simulations.  The linear theory bias discussed above and shown by the 
dashed line in Figure 2 is always positive.  Therefore, there must be a 
countervailing effect that biases $f(\Omega)$ estimates in the opposite 
direction on small scales.  In highly non-linear regions of the density field,
such as the cores of galaxy clusters, linear theory velocity predictions have
large errors.  However, errors caused by virial motions are uncorrelated with
the predicted velocities because these virial motions have random directions. 
Such errors add random scatter to the velocity-velocity relation, but they do
not change its slope.  In mildly non-linear regions of the density field, on
the other hand, galaxy velocities still follow coherent flows, but these flows
may no longer be accurately predicted by linear theory.  In the case of a
galaxy falling towards a large over-density, linear theory will correctly
predict the direction of motion, but it will overestimate the infall speed
because it incorrectly assumes that the over-density has grown at the linear
theory rate over the history of the universe, while in reality the
over-density grows to large amplitude only at late times when it becomes
non-linear.  In such regions, $\Delta \vv$ will be opposite in sign to
$\vpred$, causing an anti-correlation between the two quantities.  The 
opposite happens in under-dense regions, but since fewer galaxies reside in
these regions and the velocity errors are smaller in magnitude, the net effect 
is still an anti-correlation between $\Delta \vv$ and $\vpred$.

In order to show how these different effects come into play, we adopt a fluid
dynamics description and divide an individual galaxy's velocity into a mean 
flow $\vavg$ and a random ``thermal'' velocity $\vsig$, so that $\vtrue =
\vavg + \vsig$.  Here $\vavg(\vx)$ is the average velocity of galaxies at 
spatial position $\vx$, and therefore $\langle\vsig \cdot \vavg\rangle =$ 0 
by definition.  Let $\vlin$ denote the velocity predicted in linear theory from the
{\it unsmoothed} density field (eq.~\ref{eqn:vlin}).  
Equation~(\ref{eqn:dv}) applies to the case where the velocity field
is exactly linear, $\vtrue = \vlin$, but more generally,
\begin{eqnarray}
\vtrue & = & \vavg + \vsig \nonumber \\
       & = & \vpred + (\vlin - \vpred) + (\vavg - \vlin) + \vsig ~,
\label{eqn:dv3a}
\end{eqnarray}
and, therefore,
\begin{equation}
\Delta \vv = \vtrue - \vpred = (\vlin - \vpred) + (\vavg - \vlin) + \vsig ~.
\label{eqn:dv3b}
\end{equation}
This equation shows the three possible sources of error in the smoothed
linear theory prediction of galaxy velocities.  The first term represents
the effect caused by comparing a smoothed quantity with an unsmoothed quantity
in linear theory and is given by equation~(\ref{eqn:dv}).  The second term 
represents the inadequacy of using a linear theory velocity estimator in regions
where non-linear effects are important. The third term represents errors caused 
by galaxies' random thermal motions.  

As shown in equation~(\ref{eqn:slope}), the bias 
in $f(\Omega)$ depends on the cross-correlation of these errors with $\vpred$,
\begin{equation}
\langle\Delta \vv \cdot \vpred\rangle = \langle(\vlin - \vpred) \cdot 
			    \vpred\rangle + 
                            \langle(\vavg - \vlin) \cdot \vpred\rangle + 
                            \langle\vsig \cdot \vpred\rangle ~.
\label{eqn:dvv3}
\end{equation} 
The first term is positive and causes an overestimate of $f(\Omega)$ for
nearly all smoothing functions.  Our calculation of this effect via 
equation~(\ref{eqn:dvvpred}) shows that it is zero for no smoothing and 
increases monotonically with smoothing scale.  We have argued above that the 
second term is generally negative and causes an underestimate of $f(\Omega)$. 
Since this effect arises from the non-linearity in the density field, it should
dominate on small scales and vanish with increased smoothing of the density 
field. Finally, the third term is equal to zero because the thermal velocities
have random directions.  A combination of the first two terms 
of equation~(\ref{eqn:dvv3}) explains the scale dependence of $f(\Omega)$ 
estimates in Figure 2.  For large smoothing of the density field, the first 
term dominates and we overestimate $f(\Omega)$, whereas for small smoothing 
the second term dominates and we underestimate $f(\Omega)$.
The estimate of $f(\Omega)$ is unbiased at the smoothing scale where these
two effects cancel, but this scale should itself depend on the specifics of
the underlying cosmological model.  The numerical results in Figure 2 confirm
this prediction: the $f(\Omega)$ estimate is unbiased at $R_s=5\hmpc$ in the
$\Omega=0.2$ model (with $\sigma_8=1.44$; squares) and at $R_s=4\hmpc$ in the
$\Omega=1$ model (with $\sigma_8=0.55$; circles).
The smoothing scale for unbiased estimates could also depend on the assumed
relation between galaxies and mass, a point we will investigate in future work.
It is therefore, not possible to remove this bias simply by choosing the
right smoothing scale in a model-independent way. 

If we had adopted a higher-order perturbative expansion for predicting
velocities from the smoothed density field, then equation~(\ref{eqn:dvv3})
would still hold with $\vlin$ replaced by $\vper$, the perturbative
prediction in the absence of smoothing.  The first term on the right hand side
would still be positive, since some Fourier modes would contribute to both
($\vper-\vpred$) and $\vpred$.  The second term could be positive or negative
depending on the approximation and the smoothing scale.  However, while a
higher-order approximation might reduce the magnitude of the second term
relative to the linear approximation, it would not necessarily reduce the net
bias in $f(\Omega)$, since this depends on the relative magnitude and sign of
the first two terms.

\section{Discussion}

The implications of our results for existing estimates of $f(\Omega)$
(or, more generally, of $\beta$) are probably limited. As already mentioned, 
density-density comparisons via POTENT are not influenced by the effects 
discussed here, because they compare density and velocity divergence fields 
smoothed at the same scale.  The analysis of Davis et al. (1996), a
mode-by-mode comparison of density and velocity fields, is also not affected, 
since the two fields are again compared at the same effective ``smoothing''.  
If the observed velocities are unsmoothed, a comparison in which velocities are 
predicted using a truncated spherical harmonic expansion of the density field 
(e.g., \cite{blakeslee99}) may behave rather like our sharp $\vk$-space 
filter analysis (open symbols in Figure 2), since for a Gaussian field the 
different spherical harmonic components are statistically uncorrelated 
(A. Nusser, private communication; \cite{fisher95}).  Among recent
velocity-velocity studies, our procedure here is closest to the VELMOD analyses of
Willick et al.\ (1997) and Willick \& Strauss (1998), who
used a $3\hmpc$ Gaussian filter to compute the predicted velocity field.
These authors chose their smoothing scale partly on the basis of tests on
N-body mock catalogs, and our results in Figure~2 suggest that biases 
in $f(\Omega)$ should indeed be small for this smoothing.  However, we have 
shown that the disappearance of the bias in $f(\Omega)$ at this smoothing
scale occurs because of a cancellation between positive and negative biases,
and that the scale at which this cancellation occurs depends at least to some
degree on the underlying cosmological model.  As improvements in observational
data reduce the statistical uncertainties in peculiar velocity data, control
of the systematic uncertainties that arise from comparing smoothed velocity
predictions to unsmoothed data will become essential to obtaining robust
estimates of the density parameter.

\acknowledgments 
We thank Adi Nusser and Michael Strauss for helpful input and comments
and Marc Davis and Jeff Willick for comments on the draft manuscript.
This work was supported by NSF grant AST-9802568.  VKN acknowledges
support by the Presidential Fellowship from the Graduate School of The
Ohio State University.
 
\vfill\eject

\vfill\eject

\begin{figure}
\centerline{
\epsfxsize=6.0truein
\epsfbox[18 142 591 705]{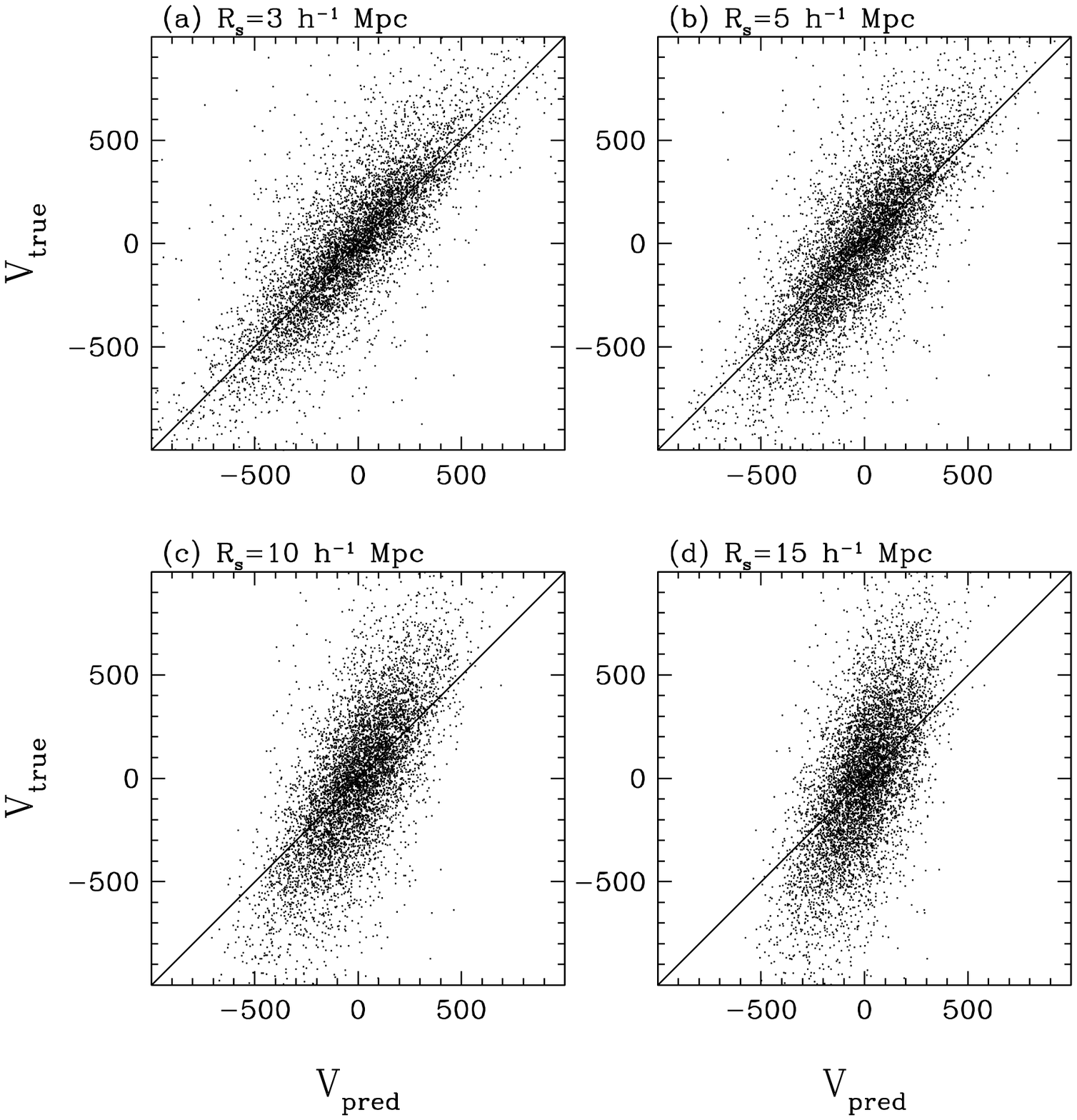}
}
\caption{
True velocities of particles from an $\Omega=1$, CDM, cosmological 
N-body simulation compared to the velocities predicted by linear theory
from the mass density field after it is smoothed with Gaussian filters of
radius $R_s=3,5,10$ and $15 h^{-1}$Mpc (panels a,b,c and d, respectively). 
The points show one Cartesian component of the particles' velocities. 
Solid lines show the relation $\vtrue=\vpred$.
} 
\end{figure}

\begin{figure}
\centerline{
\epsfxsize=7.0truein
\epsfbox[18 200 591 605]{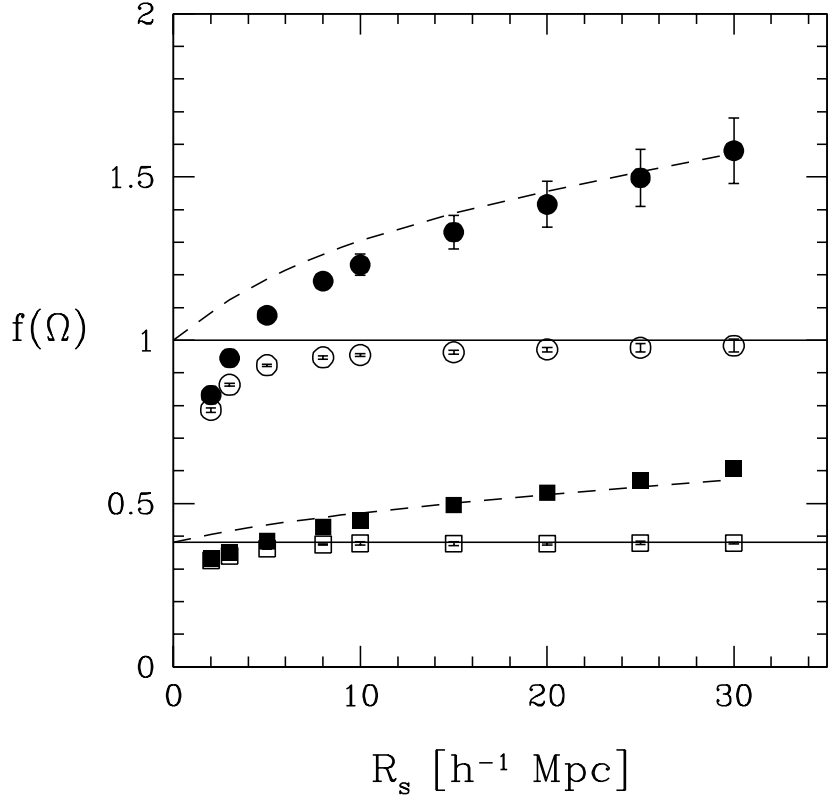}
}
\caption{
Estimates of $f(\Omega)$ from the slope of the relation between true galaxy
velocities and velocities predicted by linear theory from the smoothed density
field, as a function of the smoothing radius $R_s$, for CDM models with 
$\Omega=1$ and $0.2$ (circles and squares, respectively).  Points represent
the mean result of four simulations of each model, and error bars show the 
uncertainty in the mean derived from the dispersion among the simulations. 
Filled symbols show the estimated $f(\Omega)$ when the density field is
smoothed with a Gaussian filter of radius $R_s$.  Open symbols show the 
estimated $f(\Omega)$ when the density field is smoothed with a sharp low-pass
$\vk$-space filter (with a cut at $k_{\mathrm{cut}}$), where $R_s$ is the
radius of a Gaussian filter that falls to half its peak value at 
$k=k_{\mathrm{cut}}$.  Dashed lines show the linear theory prediction of the
bias in the estimates of $f(\Omega)$ (eqs.~\ref{eqn:slope} and 
~\ref{eqn:dvvpred}) from comparing smoothed velocity predictions to unsmoothed
velocity measurements.  Solid lines show the true values of $f(\Omega)$.
} 
\end{figure}

\end{document}